# Effect of Piezoelectric Polarization on Phonon Relaxation Rates in Binary Wurtzite Nitrides

Bijay Kumar Sahoo
Department of Physics, N. I.T, Raipur, India.
Email: bksahoo.phy@nitrr.ac.in

**Abstract**

The piezoelectric (PZ) polarization effect enhances the phonon group velocity in wurtzite nitrides. This enhancement influences the phonon relaxation rates. We calculate the modified phonon relaxation rates in binary wurtzite nitrides ( GaN, AlN and InN) by considering process like umklapp process, point defect, dislocation, boundary and phonon-electron scattering. The result will be useful to study the effect of PZ polarization on thermal conductivity of binary wurtzite nitrides (GaN, AlN and InN).

**Key words**:  A. binary wurtzite nitrides, B. piezoelectric effect, C. phonon relaxation rates.

1.  INTRODUCTION

The wurtzite nitride semiconductors (GaN, AlN, InN and their alloys) are promising materials for the next generation of high-power optoelectronics devices [1, 2]. One of the important issues in further development of wurtzite nitride is self-heating. Self-heating strongly affects the performance of the device [3, 4]. Recently, wurtzite nitride LEDs suffer from the problem of efficiency droop which resulted from the piezoelectric effect [5]. The material in the active layer and the substrate material generally determine the thermal resistance of the device structure. Thus, it is important to know the accurate value of the thermal conductivity, $\kappa$ of corresponding material.

S. Adachi has reported the experimental $\kappa$  value 1.95 W/cm K for GaN films [6]. Some samples with lower dislocation density the thermal conductivity value reaches $\kappa = 2.1$ W/ cm.K. The theoretical studies of the thermal conductivity of bulk GaN, conducted by Zou and co-workers [7], on the basis of the Callaway - Klemens approach [8, 9], have reported $\kappa$ value from 3.36 to 5.40 W/cm K by taking two sets of material parameters for GaN. They have also demonstrated the effect of impurities and dislocations on thermal conductivity [10]. The piezoelectric polarization effect on thermal conductivity, so far in our knowledge has not been considered. As Nitride wurtzites are strong piezoelectric polarization materials [11], the effect should be taken to study the change in the thermal conductivity.  The effect may bring theoretical values of thermal conductivity close to experimental value and can play a key role in determining the thermal budget at the active region of the nitride devices.

In our earlier work, we had calculated the group velocity of the phonons by taking piezoelectric polarization property and found that the effect enhances the phonon group velocity [12, 13]. Using modified phonon group velocity, calculations have been performed for Debye temperature and frequency for both binary and ternary wurtzite nitrides [13]. In this work, we have calculated the effect of the enhanced group velocity on phonon relaxation rates by considering different relaxation process such as Umklapp process, point defect, phonon boundary scattering, dislocation scattering and phonon-electron scattering in binary wurtzite nitrides (GaN, AlN and InN).

## 2. PHONON RELAXATION RATES

According to Callaway model [8, 9], the thermal conductivity $k$ is given as

$$k(T) = \left(\frac{k_B}{\hbar}\right)^3 \frac{k_B}{2\pi^2 v} T^3 \int_0^{\theta_D/T} \frac{\tau_c\, x^4\, e^x}{(e^x - 1)^2}\, dx, \qquad (1)$$

where $\theta_D$ is the Debye temperature, $k_B$ is the Boltzmann constant, $v$ is the phonon group velocity, $x = \hbar\omega/k_B T$, $\omega$ is the phonon frequency, and $\tau_c$ is the combined phonon relaxation time. The two important parameters influencing $k$ are the phonon group velocity $v$ and the combined relaxation time $\tau_c$ [14]. The polarization-averaged phonon group velocity along a specified crystallographic direction is

$$v = \left[\frac{1}{3}\left(\frac{1}{v_{T1}} + \frac{1}{v_{T2}} + \frac{1}{v_L}\right)\right]^{-1}, \qquad (2)$$

$v_L$ and $v_{T1,2}$ are the longitudinal and transverse phonon velocities, respectively. The longitudinal velocity is defined as $v_L = (C_{33}/\rho)^{1/2}$ and without piezoelectric polarization effect the transverse velocity of phonon is $v_T = (C_{44}/\rho)^{1/2}$. When piezoelectric effect is considered the transverse velocity becomes [12] $v_{T,P} = [(C_{44} + (e_{15}^2/\varepsilon_0\varepsilon_{11}))/\rho]^{1/2}$, $e_{15}$ is PZ polarization constant, $\varepsilon_0$ is permittivity and $\varepsilon_{11}$ is dielectric constant of the material along $x$ direction. Along [0001] direction, two transverse branches are degenerate and have the same velocity given by $v_{T1} = v_{T2} = v_{T,P}$. Here $C_{ij}$ are the elastic constants of the crystal [15]. The material parameters of GaN, AlN and InN are given in table1.

**Table1.** Material parameters and constants of GaN, AlN and InN [15, 18, 13].

| Parameter | GaN | AlN | InN |
|---|---|---|---|
| $a_0$ (in A⁰) | 3.112 | 3.189 | 3.54 |
| $c_0$ (in A⁰) | 4.982 | 5.185 | 3.54 |
| $e_{15}$ (C/m²) | -0.48 | -0.40 | -0.40 |
| $\varepsilon_{11}$ (= $\varepsilon_{xx}$) | 9.0 | 9.5 | 15.3 |
| $C_{33}$ (in GPa) | 373 | 405 | 224 |
| $C_{44}$ | 116 | 95 | 48 |
| $v$ (m/s) (without PZ effect) | 4746 | 6999 | 3234 |
| $v$ (m/s) (with PZ effect) | 4768 | 7067 | 329 |
| $\omega_D$ (without PZ effect) | 1.707818 x 10¹³ | 2.041050 x 10¹³ | 1.374585 x 10¹³ |
| $\omega_D$ (with PZ effect) | 1.715530 x 10¹³ | 2.060766 x 10¹³ | 1.402005 x 10¹³ |
| $\rho$ (Kg/m³) | 6150 | 3257 | 6810 |
| Deformation Potential (eV) | 8.3 | 9.5 | 7.1 |

The major phonon relaxation processes which limit $\kappa$ of a solid are : three-phonon umklapp ($\tau_c$), mass-difference ($\tau_m$), dislocation ($\tau_d$), boundary ($\tau_b$), phonon-electron scatterings ($\tau_{ph-e}$). The scattering rate for umklapp processes at high temperature ( T= 300 K and above ) is given by[11]

$$\frac{1}{\tau_u} = 2\gamma^2 \frac{k_B T}{\mu V_0 \omega_D} \omega^2, \qquad (4)$$

where $\gamma$ is the Gruneisen anharmonicity parameter, $\mu = v_T^2 . \rho$ is the shear modulus, $V_0 = \frac{\sqrt{3}}{8} a^2 c$ is the volume per atom, and $\omega_D$ is the Debye frequency. The phonon relaxation on mass-difference is [10]

$$\frac{1}{\tau_m} = \frac{V_0 \Gamma \omega^4}{4 \pi v^3} \qquad (5)$$

where $\Gamma = \sum_i f_i (1 - \frac{M_i}{M})^2$ is the measure of the strength of the mass-difference scattering, $f_i$ is the fractional concentration of the impurity atoms, $M_i$ is the mass of the $i$ th impurity atom,

$\bar{M} = \sum_i f_i M_i$ is the average atomic mass. The phonon scattering rate at the core of the dislocation is proportional to the cube of the phonon frequency and is given by [11]

$$\frac{1}{\tau_d} = \eta \, N_D \, \frac{V_0^{4/3}}{v^2} \, \omega^3, \tag{6}$$

where $N_d$ is the density of the dislocation lines of all is types, and $\eta = 0.55$ is the weight factor to account for the mutual orientation of the direction of the temperature gradient and the dislocation line. The boundary scattering relaxation rate is given in the Casimir limit as

$$\frac{1}{\tau_b} = \frac{v}{L}, \tag{7}$$

where L is the dimension of the sample. At low doping levels, the relaxation time for acoustic phonons scattered by electrons can be expressed as [16]

$$\frac{1}{\tau_{ph-e}} = \frac{n_e \, \epsilon_1^2 \, \omega}{\rho \, v^2 \, k_B \, T} \sqrt{\frac{\pi \, m^* v^2}{2 \, k_B \, T}} \, exp\left(-\frac{m^* v^2}{2 \, k_B \, T}\right), \tag{8}$$

Where $n_e$ is the concentration of conduction electrons, $\epsilon_1$ is the deformation potential, $\rho$ is the mass density, and $m^*$ is the electron effective mass. The combine phonon relaxation time can be obtained by the summation of the inverse relaxation times for these scattering processes and is given by

$$\frac{1}{\tau_c} = \frac{1}{\tau_u} + \frac{1}{\tau_m} + \frac{1}{\tau_d} + \frac{1}{\tau_d} + \frac{1}{\tau_{ph-e}}. \tag{9}$$

## 3. RESULT AND DISCUSSION

The combine relaxation rate is calculated by calculating Umklapp phonon scattering, mass-difference, dislocation, boundary and phonon-electron scattering processes. The variation of phonon scattering rates with phonon frequency considering with and without PZ polarization effect are shown in figures. The material parameters used in our calculation are noted in Table 1. Fig.1 shows the variation of scattering rates for GaN. Here $\Gamma = 1.3 \times 10^{-4}$, $N_D = 1.5 \times 10^{10}$ cm$^{-2}$, $n_e = 2.6 \times 10^{18}$ cm$^{-3}$, $\epsilon_1 = 8.3$ eV, $m^* = 0.22$ m$_0$, and L= 22nm are used [17]. It can be seen that the difference between the two curves is distinguishable at low phonon frequency and as the frequency increases the two lines overlap. Hence, in GaN the effect of PZ polarization on combine scattering rate become appreciable at low phonon frequency while at high frequency the combine scattering rate are not affected by PZ polarization.

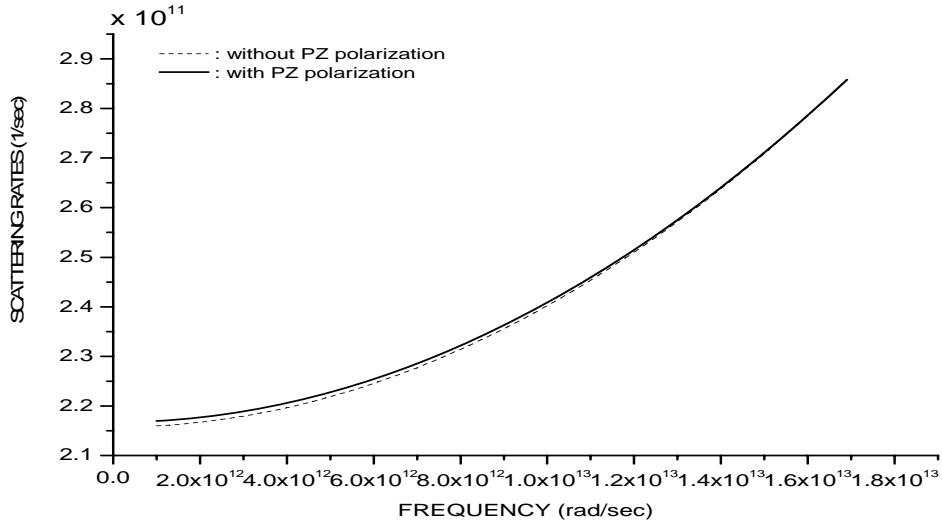

FIG.1.Combine scattering rates in GaN at T= 300K. The result is shown for the three-phonon Umklapp, mass - difference, dislocation, boundary and phonon - electron scattering.

Fig.2 shows the combine scattering rates variation with phonon frequency for with and without PZ polarization effect at room temperature. Here $\Gamma = 3.2 \times 10^{-4}$, $N_D= 1.5 \times 10^{10}$ cm$^{-2}$, $n_e= 4.1 \times 10^{18}$ cm$^{-3}$, $\epsilon_1= 9.5$ eV, $m^*= 0.3$ m$_0$, and L= 22nm are used [ 18, 19].

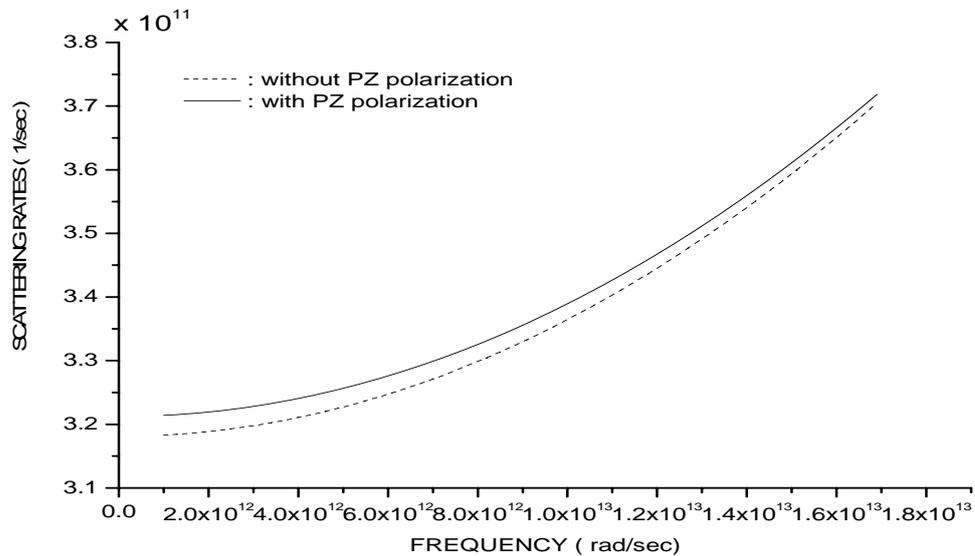

FIG.2.Combine scattering rates in AlN at T= 300K. The result is shown for the three-phonon Umklapp, mass - difference, dislocation, boundary and phonon - electron scattering.

It can be observed that at low frequency the effect of PZ polarization on combine scattering rates is higher than at high frequency. Thus, the combine scattering rate is most affected by PZ polarization effect in case of AlN. Fig.3 shows the combine phonon relaxation rates with phonon frequency. Here $\Gamma = 3.2 \times 10^{-4}$, $N_D = 5 \times 10^{10}$ cm$^{-2}$, $n_e = 3.3 \times 10^{18}$ cm$^{-3}$, $\epsilon_1 = 7.1$ eV, $m^* = 0.14 m_0$, and L= 22 nm are used[20, 19].

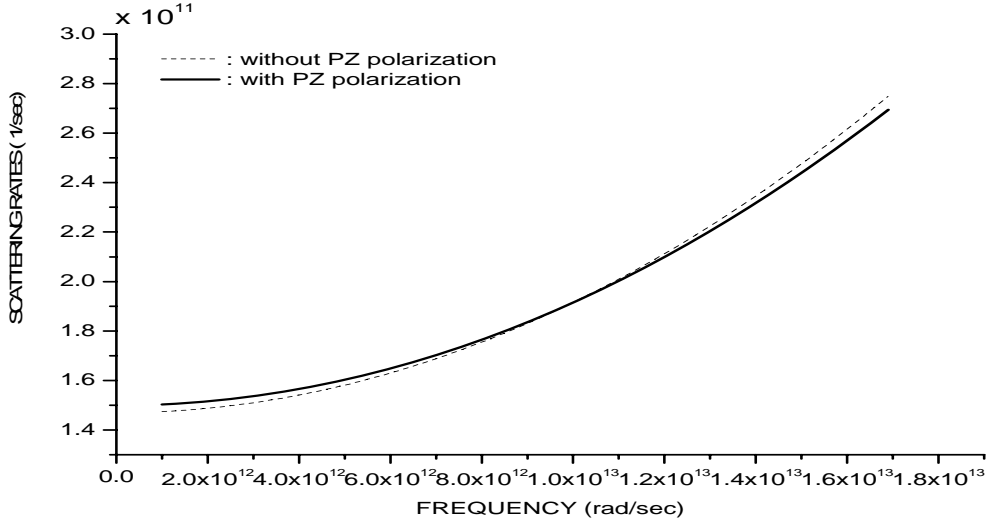

FIG.3. Combine scattering rates in InN at T= 300K. The result is shown for the three-phonon Umklapp, mass- difference, dislocation, boundary and phonon- electron scattering.

From Fig.3, it can seen that both at low and high phonon frequency, relaxation rates are different while at frequency $1.0 \times 10^{13}$ (rad/sec) there is a crossover of the both graphs. Thus, it can be said that the combine relaxation rates are affected by PZ polarization effect both at high and low frequency in InN. This result is a new findings in this work.

## 4. CONCLUSIONS

We have investigated in theory the effect of piezoelectric polarization on phonon relaxation rates in binary wurtzite nitrides. The strongest variation of the relaxation rates due to PZ polarization is found in case of AlN. The result can be used for evaluating the piezoelectric polarization effect on thermal conductivity of binary wurtzite nitrides at room temperature.

### ACKNOWLEDGEMENT
The author thanks Professor Dr. S. M. Saini, Department of Physics, National Institute Technology, Raipur, India for fruitful discussions.